\documentclass[preprint,12pt]{elsarticle}%\documentclass{svjour2} 
\usepackage{amsmath,amssymb,amsbsy,amsfonts,latexsym}
\usepackage{graphicx}
\newcommand{\be}{\begin{equation}}    % for lazy typers
\newcommand{\ee}{\end{equation}}
\newcommand{\ba}{\begin{eqnarray}}
\newcommand{\ea}{\end{eqnarray}}
\newcommand{\bv}{\boldsymbol}
\catcode`@=11
\newcommand\eqalign[1]{\null\,\vcenter{\openup\jot\m@th
    \ialign{\strut\hfil$\displaystyle{##}$&$\displaystyle{{}##}$\hfil
        \crcr#1\crcr}}\,}

\journal{New Astronomy}

\begin{document}

\begin{frontmatter}

\title{Normal forms for the epicyclic approximations of the perturbed Kepler problem}
\author{Giuseppe Pucacco}
             
             \address{
              Physics Department, University of Rome ``Tor Vergata" and
INFN - Sez. Roma II
              {\rm e-mail:}{pucacco@roma2.infn.it}}

\date{Received: date / Accepted: date}

\begin{abstract}
We compute the normal forms for the Hamiltonian leading to the epicyclic approximations of the (perturbed) Kepler problem in the plane. The Hamiltonian setting corresponds to the dynamics in the Hill synodic system where, by means of the tidal expansion of the potential,  the equations of motion take the form of perturbed harmonic oscillators in a rotating frame. In the unperturbed, purely Keplerian case, the post-epicyclic solutions produced with the normal form coincide with those obtained from the expansion of the solution of the Kepler equation. In all cases where the perturbed problem can be cast in autonomous form, the solution is easily obtained as a perturbation series. The generalization to the spatial problem and/or the non-autonomous case is straightforward.

\end{abstract}


\end{frontmatter}

\section{Introduction}

The epicyclic approximation of the Kepler problem consists in the superposition of the motion of a guiding center along a circle and a  motion along an ellipse of axis ratio 1:2. The radius of the main circle coincides with the semi-major axis $a$ of the exact Keplerian orbit. The semi-minor axis of the ellipse is a factor $e$ smaller, with $e$ the eccentricity of the orbit. 

The power of the epicyclic formulation lays in being explicitly expressed in terms of the mean anomaly, therefore of time. This property is preserved by the `post-epicyclic' approximation, namely the superposition of higher harmonics of the mean motion providing a series of periodic deformations of the fundamental ellipse. 

The epicyclic and the post-epicyclic approximation can be considered as a perturbation expansion in which the circular orbit is the zero-order solution, the epicycle is a first-order term and the post-epicycles are higher-order corrections. The eccentricity is the natural small perturbation parameter. A standard way to get the post-epicyclic approximation of the Keplerian orbits is through the series expansion of the Kepler equation as given by Laplace. There is an alternative approach to get solutions in the form of series by casting the problem in the perturbative setting of the Hill (or synodic) system: the dynamical problem is equivalent to that of an oscillator in a uniformly rotating frame. The linear harmonic approximation gives the standard epicyclic solution, nonlinear coupling terms give rise to higher harmonics. 

There has recently been a revival of interest in the post-epicyclic approximations (the LISA experiment (Durandhar et al., 2005; Sweetser, 2005; Pucacco et al., 2010), coorbital motion of Saturn satellites (Gurfil and Kasdin, 2003), the $J_2$ problem (Kasdin et al., 2005), etc.) due to the fact that the Hill frame provides a quite accurate description of perturbations of low eccentricity orbits, just by solving a simple quasi-linear mechanical system. It can be shown that, up to second order, the post-epicyclic approximation and the perturbed solution in the Hill system coincide (Rajesh Nayak et al., 2006). A natural question is up to which order, if any, this equivalence persist. In this regard, we recall here a classic result of celestial mechanics: the unperturbed Keplerian expansion is correctly reconstructed at arbitrary order by solving the equations of motion in the Hill system (Brouwer and Clemence, 1961). Although the same technique could be employed also to treat the perturbed case, we prefer to take a different approach. Aim of this paper is to investigate the perturbed Keplerian problem by constructing a Hamiltonian normal form and to get from it perturbation series solutions: we verify that the expansions for the unperturbed Keplerian orbits are correctly reconstructed. Then we proceed to solve samples of perturbed Keplerian problems.

The plan of the paper is as follows: in Section 2 we recall the relation between the series solutions of the Kepler equation, the epicyclic formulas and the construction of the Hill system for the equations of motion; in Section 3 we introduce the Hamiltonian normalization setting that we use to cope with the solution of the perturbed system in the synodic frame; in Section 4 we construct the normal form and compare its prediction with those obtained from the Kepler equation; in Section  5 we apply this formalism to the study of some examples of perturbed Keplerian problems; in Section 6 we give our conclusions.

\section{Epicyclic approximations of the Kepler problem}

\begin{figure}[h!]
\centering
\includegraphics[width=0.69\columnwidth]{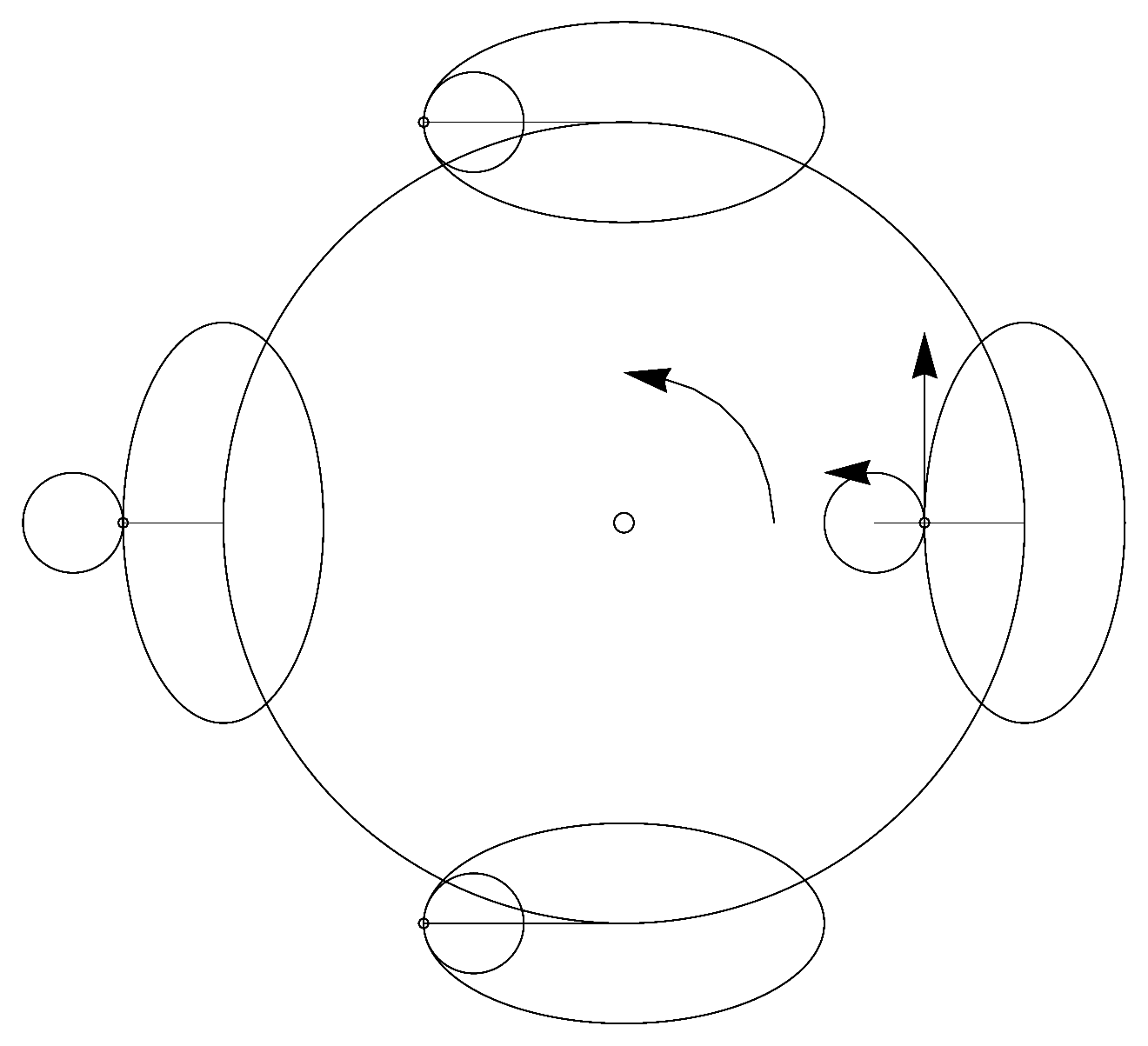}
\caption{Epicycle representation of a Keplerian orbit.}
\label{hcw}
\end{figure}

 We consider a Keplerian orbit of semi-major axis $a$ and eccentricity $e$ on the plane $X,Y$ in the field generated by the mass $M$. A simple way to get the epicyclic approximation is by inserting the series expansion of the solution of the Kepler equation
\begin{equation}\label{KEQ}
u - e \sin u = n t ,
\end{equation}
into the parametric form
\begin{eqnarray}
X&=&a (\cos u-e), \label{eqx}\\
Y&=&a \sqrt{1-e^{2}}\sin u \label{eqy}
\end{eqnarray}
of the orbit. In these expressions $u$ is the eccentric anomaly, the mean motion is
\be \label{KMM} n = \sqrt{\frac{GM}{a^{3}}}, \ee
and the product $n t$ is referred to as the mean anomaly also denoted with $\ell$ in the following. The series expansion  in the eccentricity of the solution of the Kepler equation (\ref{KEQ}) is (Boccaletti and Pucacco, 1999)
\be \label{KEX4}
u = n t + e \sin n t + \frac12 e^2 \sin 2  n t + e^3 \left(\frac38 \sin 3  n t - \frac18 \sin  n t \right) + ...\ee
Inserting this into (\ref{eqx}) and (\ref{eqy}) and expanding in series of the eccentricity gives two Fourier series that can be referred to as {\it epicyclic expansions} of the Keplerian orbit: 
\begin{eqnarray}
X&=&a \left[\cos  n t - \frac{e}2 \left(3 - \cos 2  n t \right) +...\right], \label{fx1}\\
Y&=&a \left[\sin  n t + \frac{e}2 \sin 2  n t +...\right].\label{fy1}
\end{eqnarray}
In Fig.\ref{hcw}, which is inspired by a picture appearing in Sweetser (2005), we can see the role of the different terms in the expansions (\ref{fx1}--\ref{fy1}): the uniform motion on the great circle is given by the trigonometric terms in $nt$ of amplitude $a$; the motion on the ellipse is produced by the terms in $2nt$ of amplitude $ae/2$ displaced along $X$ by the constant amount $-3ae/2$. The superposition of these two motions provides the first-order approximation of the Keplerian orbit.

\begin{figure}[h!]
\centering
\includegraphics[width=0.89\columnwidth]{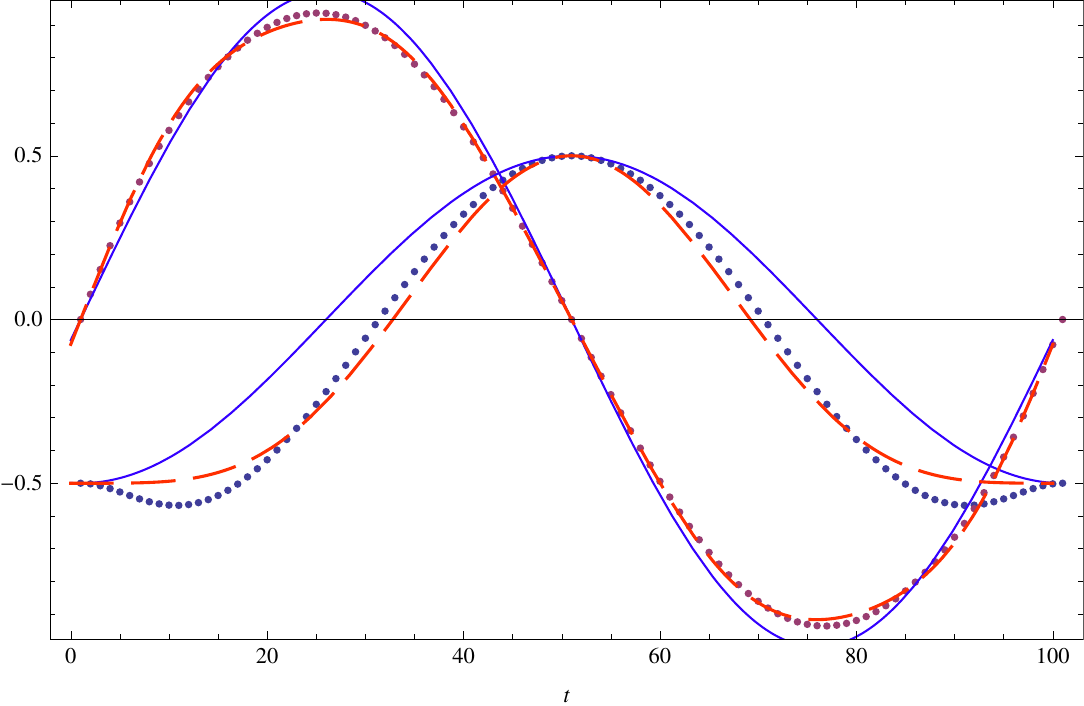}
\caption{Comparison between the exact Keplerian solution (dots) of an orbit with $a = 1, \; e = 0.5$ and the first two terms in the post-epicyclic expansion: the continuous curves represent the epicyclic (first-order) terms; the dashed curves include the contribution of the second-order terms.}
\label{epiciclic}
\end{figure}

For our needs it is more convenient to express these expansions in the frame rotating with the mean motion ({\it synodic} frame) which turns out to be ideally suited to this perturbative approach. We make the change of coordinates
\ba
x &=& X \cos  n t + Y \sin n t - a, \label{eqxr}\\
y &=& Y \cos  n t - X \sin n t, \label{eqyr}\ea
so that
\begin{eqnarray}
x&=&ae \left[-\cos  n t - \frac{e}2 \left(1 - \cos 2  n t \right) + 
                                           \frac38 e^2 \left(\cos 3  n t - \cos  n t \right) +...\right], \label{esx1}\\
y&=&ae \left[2 \sin  n t + \frac{e}4 \sin 2  n t + 
                                           \frac18 e^2 \left(\frac73 \sin 3  n t - 3 \sin  n t \right) +...\right].\label{esy1}
\end{eqnarray}
The first-order terms in these expansions provide the properly called epicyclic solution: an ellipse with the semi-minor axis of length $ae$ along the (radial) $x$-axis and the semi-major axis of length $2ae$ along the (negative, tangential) $y$-axis. The higher-order terms are higher harmonic periodic corrections. In Fig.\ref{epiciclic} we see the comparison between the exact Keplerian solution and the first two terms in the expansion: an unusually high value of the eccentricity is chosen for the plots ($e = 0.5$) in order to magnify the discrepancy between the true orbit and its approximations.

An alternative route to the epicyclic approximation is that based on the solutions of the perturbed equations of motion in the Hill frame which is convenient when we have to analyze in the simplest way the geometry of low eccentricity orbits and their most relevant perturbations. The Hill reference system (see, Boccaletti and Pucacco, 1999 and Bik et al. 2007), also called Clohessy-Wiltshire system (Clohessy and Wiltshire, 1960) allows us to reduce the equations of motion to those of a set of coupled oscillators. The Hill equations properly said consist of the linear part of this system which is readily solved. Non-linear and non-autonomous terms can be added and solved perturbatively. 

The synodic reference system given by transformations (\ref{eqxr},\ref{eqyr}) is based on a circular orbit of radius $a$ around the central mass and has its origin rotating at angular velocity $n$ with the reference orbit. In the applications to a perturbed Kepler problem, it is useful to relax this assumption and the angular velocity can be arbitrary, say $\Omega$. Reverting for the time being to the three dimensional description for a clearer geometric view, we have the axes $x,y,z$ such that $x$ is directed radially opposed to the center, $y$ is in the direction tangent to the motion and $z$ is perpendicular to their plane. The new equations of motion are introduced according to the following steps:

1. Transform from the inertial barycentric system to the non-inertial synodic system: the equations of motion take the form

\begin{equation}\label{hill}
\ddot {\bv x} = - \nabla \Phi - 2 \bv \Omega \wedge \dot {\bv x}
- {\bv \Omega} \wedge \left({\bv \Omega} \wedge ({\bv x} + \bv a) \right).
\end{equation}
Here the dot denotes the derivative with respect to time, $\Phi$ is the total gravitational potential and ${\bv x} = (x,y,z), {\bv \Omega} = (0,0,\Omega), {\bv a} = (a,0,0)$. 

2. The potential is approximated according to the expansion

\begin{equation}\label{hill2}
\Phi = \Phi \vert_{\bv x=0} + 
   \frac{\partial \Phi}{\partial x_{\lambda}} \bigg\vert_{\bv x=0} x_{\lambda} 
+ \frac12 \frac{\partial^{2} \Phi}{\partial x_{\mu} \partial x_{\lambda}} \bigg\vert_{\bv x=0} x_{\mu} x_{\lambda} 
+ {\rm O} (|\bv x|^{3}),
\end{equation}
where the sum is implicit over the repeated indexes $\lambda, \mu =1,2,3$. The ``far tide'' (or Hill) approximation (Boccaletti and Pucacco, 1999) consists of retaining only the quadratic terms in (\ref{hill2}) and therefore only first order terms into the expansion of the gradient of the potential. 

Taking as reference example the unperturbed Kepler potential we have

\begin{equation}
\Phi=\Phi_{\rm K} \doteq -\frac{G M}{\sqrt{(x+a)^{2}+y^{2}+z^{2}}}
\end{equation}
so that the Keplerian far tide approximation gives

\begin{equation}
\frac{\partial\Phi_{\rm K}}{\partial x} \bigg\vert_{\bv x=0}=\frac{G M}{a^{2}}= n^{2} a, \;\;\;\;
\frac{\partial\Phi_{\rm K}}{\partial y} \bigg\vert_{\bv x=0}= \frac{\partial\Phi_{\rm K}}{\partial z} \bigg\vert_{\bv x=0} = 0 \end{equation}
and

\begin{equation}
\frac{\partial^{2} \Phi_{\rm K}}{\partial x^{2}} \bigg\vert_{\bv x=0}=-
\frac{2 G M}{a^{3}}=-2  n^{2}, \;\;\;\;
\frac{\partial^{2} \Phi_{\rm K}}{\partial y^{2}} \bigg\vert_{\bv x=0} = 
\frac{\partial^{2} \Phi_{\rm K}}{\partial z^{2}} \bigg\vert_{\bv x=0} = \frac{G M}{a^{3}}=  n^{2}, \end{equation}
all others zero. The next order of approximation ({\it octupole}) is:

\begin{eqnarray}
\frac{\partial^{3} \Phi_{\rm K}}{\partial x^{3}} \bigg\vert_{\bv x=0}&=&
\frac{6 G M}{a^{4}}=\frac{6 n^{2}}{a}, \\
\frac{\partial^{3} \Phi_{\rm K}}{\partial x \partial y^{2}} \bigg\vert_{\bv x=0}&=&
\frac{\partial^{3} \Phi_{\rm K}}{\partial x \partial z^{2}} \bigg\vert_{\bv x=0}=-
\frac{3 G M}{a^{4}}=-\frac{3 n^{2}}{a}, 
\end{eqnarray}
all others zero. By using these in (\ref{hill}), with $\Omega = n$, we get

\begin{eqnarray}
\ddot x &-& 2  n \dot y - 3  n^{2} x+ \frac{3 n^{2}}{2a} (2 x^{2} - y^{2} - z^{2}) =  f_x (\bv x, \bv \varepsilon),\nonumber \\
\ddot y &+& 2  n \dot x  - \frac{3 n^{2}}{a} xy =  f_y (\bv x, \bv \varepsilon), \label{hilleq}\\
\ddot z &+&  n^{2} z - \frac{3 n^{2}}{a} xz =  f_z (\bv x, \bv \varepsilon) \nonumber
\end{eqnarray}
where in the vector $\bv f (\bv x, \bv \varepsilon)$ we include the higher-order terms produced by the expansion (\ref{hill2}) and, in  general, {\it autonomous} perturbations. We  
assume that any other perturbing force can be derived by an analogous series expansion of the corresponding potential and added to this system. Each component, coherently with a perturbative approach, will be a power series in the coordinates with suitably `small' coefficients ${\bv \varepsilon} = (\varepsilon_x, \varepsilon_y, \varepsilon_z)$.  

In these equations the second order approximation of the perturbed Keplerian motion is explicitly given. Coherently with the above treatment, we will limit the analysis to the motion on the plane and will drop the third equation for the $z$ oscillation. This does not affect too much the generality of the approach and can be easily included if desired. 

The perturbation vector, in its simplest form, consists of constant and linear (in $x$) terms of the form $ \varepsilon_0 + \varepsilon_1 x + \dots $: no terms in $y$ are allowed if we want preserve the autonomous nature of the perturbation. This gives explicit small changes already at zero-order. These and possibly higher-order perturbing terms propagate in the full solutions through the non-linear coupling with the other components of the field expansion. The presence of constant terms in the perturbation suggests to generalize the Hill equations by not specifying in advance the angular velocity and leaving it as the free parameter $\Omega$ that will typically be a small perturbation of the Keplerian value (\ref{KMM}). Rather we leave $a$ as the fixed radius of the reference circular orbit. In this way, the relevant system at first order can be written in the form

\begin{eqnarray}\label{eq2_123}
\ddot {x} &-& 2 \Omega \dot {y} - \Omega_{\rm p}^{2} x + {\rm O}(|\bv x|^2)=0,\nonumber\\
\ddot {y} &+& 2 \Omega \dot {x}  + {\rm O}(|\bv x|^2)=0, \end{eqnarray}
where 

\be \label{Omegap} 
\Omega_{\rm p}^{2} \simeq (3+2 \varepsilon) \Omega^{2} \ee
includes the first order contribution of the perturbing field and the small parameter $\varepsilon$ is determined by the nature of this field. In this work we will consider the simple class of the form $\Phi_{\rm pert} \sim 1/r^3$, but eq.(\ref{Omegap}) holds for any third-order expansion complying with the above stated conditions. Since in general the presence of quadratic or higher-order terms in the system  (\ref{eq2_123}) makes it {\it non-integrable}, the only viable route to get some solutions is to head to a perturbative approach. The solution of the system can therefore be written in the form of a series expansion

\be\label{approx}
{\bv {x}}(t) = {\bv {x}}^{(1)} + {\bv \varepsilon} \cdot {\bv {x}}^{(2)} + \dots, 
\ee
where the value of the upper index denotes the order. 

The first-order general solution is obtained by solving the linear part of (\ref{eq2_123}). The characteristic equation of this linear system is

\be \label{CE}
\lambda^4 + \lambda^2 (4 \Omega^2 - \Omega_{\rm p}^{2}) = 0, \ee
so that one of its eingenvalues vanishes and the other provides the frequency

\be \label{NF}
\omega = \sqrt{4 \Omega^2 - \Omega_{\rm p}^{2}} \simeq \Omega (1-\varepsilon).\ee
The zero eigenvalue is associated to the neutral stability of one of the normal modes therefore producing a linear drift. Choosing initial conditions in order to remove constant and secular terms due to this drift, the solution of the linear system is:

\begin{equation}\label{zeroxyz}
x^{(1)} = -A \cos (\omega t + \phi), \hskip 1.5cm
y^{(1)} = 2 \frac{\Omega}{\omega} A \sin (\omega t + \phi). 
\end{equation}
In the purely Keplerian case, 

\be
\omega = \Omega = n,
\quad \Omega_{\rm p} = \sqrt{3} n, 
\quad A = ea, 
\quad \phi=0,\label{mm}\ee
they coincide with the first terms in (\ref{esx1},\ref{esy1}) and correspond to harmonic oscillations around the `reference' $\bv x = \bv 0$ circular orbit. In this case, the perturbative setting is provided by the observation that the radial amplitude is constrained by the apsides and therefore we can rescale the coordinates according to

\be\label{scale}
\tilde x = \frac{x}{ae}, \quad \tilde y = \frac{y}{ae}.\ee
Additional perturbations add gradually smaller terms according to the size of coupling constants like those above. Higher order terms in the solution can be obtained by plugging the series  (\ref{approx}) in the system  (\ref{eq2_123}), equating terms of the same order in $\varepsilon$ and exploiting solution (\ref{zeroxyz}). On the other hand, a more general approach to the solution of perturbation problems can be obtained with the more effective method of constructing a Hamiltonian normal form, as shown in the next section. 
From the analytical point of view it is worth mentioning that even the simple system (\ref{hilleq}) with just $\bv f (\bv {\tilde x}, \bv \varepsilon) \equiv {\bv 0}$ (with ${\bv {\tilde x}}(t) = {\bv {\tilde x}}^{(1)} + e {\bv {\tilde x}}^{(2)}$
 as a second-order approximate solution) is not integrable, unlike the original Kepler system which is clearly integrable. Is a common lesson of modern analytical mechanics that truncated solutions of non-integrable approximations can be more useful than formally exact solutions of their integrable counterparts.

\section{The Hamiltonian setting}

The equations of motion in the Hill system can be obtained by introducing a suitable Hamiltonian. Applying the approach usually adopted with Hamiltonians admitting linear terms in the momenta (Hietarinta, 1987; Pucacco and Rosquist, 2005) it can easily be proven that the planar ($z=0$) set of system (\ref{hilleq}) is generated by the Hamiltonian 

\be\label{HH}
{\cal H} (\bv p, \bv x) =\frac12 (p_x + \Omega y)^2 + \frac12 (p_y - \Omega x)^2 - \frac12  \Omega_{\rm p}^2 x^2 +
\Phi^{(3)}, 
\ee
where $\Phi^{(3)}$ is the expansion of the potential (Keplerian and, possibly, perturbations) starting from the cubic terms. The canonical equation given by this Hamiltonian coincide with the first two equations of system (\ref{hilleq}) if we retain in the expansion of $\Phi_{\rm K}$ only the cubic terms and assume that  $\bv f (\bv x, \bv \varepsilon) = - \nabla \Phi_{pert}(\bv x, \bv \varepsilon)$.

The most effective way of investigating the solution of the canonical equation generated by (\ref{HH}) is by constructing a normal form by a suitable canonical transformation. We look for a new Hamiltonian in the new canonical variables $\bv{P,X}$, given by 
\begin{equation}\label{HK}
     {\cal K}({\bv{P,X}})=\sum_{n=0}^{\infty}{\cal K}_n ({\bv{P,X}}),
  \end{equation}
with the prescription that 
\be\label{NFD}
\{{\cal G},{\cal K}\}=0
\ee
where ${\cal G}$ is a suitable low degree polynomial in the canonical variables. When ${\cal H}_{0}$, the quadratic part of the Hamiltonian under investigation, is semi-simple, the usual choice is simply ${\cal G} = {\cal H}_{0}$. In the present case this is not possible and we have to make a different choice. In any case, in order to set up the procedure based on the Lie transform method,  ${\cal H}_0$ must be diagonalized (in these and subsequent formulas we adopt the convention of labeling the first term in the expansion with the index zero: in general, the `zero order' terms are quadratic homogeneous polynomials and terms of order {\it n} are polynomials of degree $n+2$). 

With the canonical transformation $(p_x,p_y,x,y) \longrightarrow (p_\xi,p_\eta,\xi,\eta)$ defined by

\ba
x &=& \frac1{\sqrt{\omega}} (\xi - p_\eta), \label{T1}\\
y &=& \frac{2 \Omega}{\omega^{3/2}} (p_\xi - \eta),  \label{T2}\\
p_x  + \Omega y &=& \sqrt{\omega} p_\xi,\\
p_y - \Omega x &=& \frac1{\sqrt{\omega}} \left( \frac{4 \Omega^{2} - \omega^2}{2 \Omega} p_\eta - 2 \Omega \xi \right),  \label{T4}\ea
we get  

\be\label{Hzero}
{\cal H}_{0} = \frac12 \omega (p_\xi^2 + \xi^2) + \frac12 {\kappa} p_\eta^2
\ee
where
\be\label{kappa}
\kappa = \frac{ \omega (\omega^2 - 4 \Omega^{2})}{4 \Omega^{2}}
            =- \frac{ \omega \Omega_{\rm p}^{2}}{\omega^2 + \Omega_{\rm p}^{2}}.
\ee
This quadratic part is of semi-simple+nilpotent type: this case has already been studied (Broer at al., 1993; Palacian and Yanguas, 2000) and it appears that a natural choice to work out the normalization is to extend only the partial energy

\be\label{GG}
{\cal G} = \frac12 (p_\xi^2 + \xi^2),\ee
which is an obvious integral of motion of ${\cal H}_0$, to the whole normal form ${\cal K}$. 

In the process of normalization based on the Lie transform, new coordinates $\bv{P,X}$ result from the canonical transformation
  \be\label{TNFD}
  ({\bv{p_\xi,\xi}}) = T_{\chi} (\bv{P,X}).\ee
Considering a generating function $\chi$, the Lie transform operator $T_{\chi}$ is defined by (Boccaletti and Pucacco, 1999)
\begin{equation}\label{eqn:OperD-F}
    T_{\chi} \equiv \sum_{k=0}^{\infty} M_k
\end{equation}
where
\be M_0 = 1, \quad M_k = \sum_{j=1}^k \frac{j}{k} L_{\chi_j} M_{k-j}.\ee
The functions $\chi_k$ are the coefficients in the expansion of the generating function of the canonical transformation and the linear differential operator $L_{\chi}$ is defined through the Poisson bracket, $L_{\chi}(\cdot)=\{{\chi},\cdot\}$.

The terms in the new Hamiltonian are determined through the recursive set of linear partial differential equations (Giorgilli, 2002)
\be\label{EHK} \begin{array}{ll}
    &{\cal K}_0={\cal H}_0 ,\\ \\
    &{\cal K}_1={\cal H}_1+M_{1}{\cal G} =  {\cal H}_1+L_{\chi_1}{\cal G},\\ \\
    &{\cal K}_2={\cal H}_2+M_{1}{\cal H}_1+M_{2}{\cal G} =  
               {\cal H}_2+L_{\chi_1}{\cal H}_1 + \frac12 L^{2}_{\chi_1}{\cal H}_0+L_{\chi_2}{\cal G} ,\\ \\
    &\quad \;\; \vdots \\ \\
    &{\cal K}_k= {\cal H}_k  +\sum_{j=1}^{k-1}M_{k-j}{\cal H}_j +M_{k-1}{\cal H}_0+L_{\chi_k} {\cal G},
\end{array}\ee
where $L^{2}_{\chi}(\cdot)=\{{\chi},\{{\chi},\cdot\}\}$. `Solving' the equation at the $k$-th step consists of a twofold task: to find ${\cal K}_{k}$ {\it and} $\chi_k$. Once a given `target' order $N = k_{\rm max}$ has been reached the process is stopped and two truncated series 

\ba
{\chi}^{(k_{\rm max})} &=& \sum_{k=1}^{N}\chi_k ({\bv{P,X}}), \label{chin} \\
{\cal K}^{(k_{\rm max})} &=& \sum_{k=0}^{N}{\cal K}_k ({\bv{P,X}}), \label{Kn}
\ea
are constructed. The new Hamiltonian ${\cal K}^{(k_{\rm max})}$  is integrable because, by construction, admits an additional integral of motion. If we are able to solve canonical equations, we get the flow $(\bv{P}(t),\bv{X}(t))$ and we can use the generating function ${\chi}^{(k_{\rm max})}$ to construct the transform (\ref{TNFD}) and the transformation rules (\ref{T1}--\ref{T4}) to obtain the flow in terms of the initial `physical' variables  $({\bv{p}(t),\bv{x}(t)})$. Also these expressions will be series truncated at the same order as the generating function.
	
In the following section we apply this theory to the Hamiltonian of the Keplerian problem in the Hill frame by explicitly constructing the series ${\bv x}(t) = {\bv x}^{(1)} + e {\bv x}^{(2)} + ...$ showing that they coincide with that obtained by expanding the series solution of the Kepler equation. Later on we outline the general approach to cope with autonomous perturbations.

\section{Normalization and construction of the generating function for the unperturbed Kepler problem}

We have explicitly constructed the two truncated series (\ref{chin}) and (\ref{Kn}) in the Keplerian case up to $k_{\rm max}=4$, namely up to and including terms of degree 6 in the canonical variables. We start by expanding the Kepler potential up to order 6, giving the functions

\ba
{\cal H}_1 &=& \frac{n}a \left(x^3-\frac{3 x y^2}{2} \right),\\
{\cal H}_2 &=& \frac{n}{a^2} \left(-x^4+3 x^2 y^2-\frac{3 y^4}{8} \right),\\
{\cal H}_3 &=& \frac{n}{a^3} \left(x^5-5 x^3 y^2+\frac{15 x y^4}{8} \right),\\
{\cal H}_4 &=& \frac{n}{a^4} \left(-x^6+\frac{15 x^4 y^2}{2}-\frac{45 x^2 y^4}{8}+\frac{5 y^6}{16} \right).
\ea
These expressions are written with a small abuse of notation with respect to the previous section, where the higher-order terms of the perturbed Hamiltonian are assumed to be already prepared for the normalization and should be expressed in terms of the new variables introduced by  (\ref{T1}--\ref{T4}). However, it is more simple and natural to use the old physical coordinates. Moreover, the Hamiltonian is rescaled dividing by the mean motion so that the new time associated to the transformed Hamiltonian is the mean anomaly $\ell = n t$. Using (\ref{kappa}) and the first two equalities of (\ref{mm}), the unperturbed diagonal part is therefore
\be\label{Hzerokep}
{\cal H}_{0} = \frac12 (p_\xi^2 + \xi^2) - \frac38 p_\eta^2.
\ee
A first notable outcome of the normalization procedure is that, up to $N=k_{\rm max}=4$, all the perturbing terms in the Hamiltonian can be removed and therefore the normal form up to this order is simply

\be\label{Kzero}
{\cal K} \equiv 
{\cal K}_{0} = \frac12 (P_X^2 + X^2) - \frac38 P_Y^2,  
\ee
with ${\cal K}_{j}=0$ for $ 1\le j\le4.$ We may conjecture that this is true for the Keplerian expansion at all orders but, anyway, we can imagine that this property is in general lost as soon as one introduces an arbitrary extra perturbation. 

The extremely simple structure of the normal form is counterbalanced by the generating function which, in view of the its role of eliminating a huge number of terms at each step, is quite cumbersome. We limit ourselves to give the explicit expressions up to order 4 which will be enough to establish our result.  We get

\begin{eqnarray}
a n \chi_1 &=& -X^2 Y+\frac{8 Y^3}{3}+X^2 P_X-6 Y^2 P_X+5 Y P_X^2 \nonumber \\
          && -\frac{4 P_X^3}{3}+3 X Y P_Y-\frac{9}{4} X P_X P_Y-\frac{4 Y P_Y^2}{3}+\frac{3}{4} P_X P_Y^2, \label{chi1}
\end{eqnarray}

\begin{eqnarray*}
a^2 n^2 \chi_2 &=& \frac{3 X^3 Y}{2}-6 X Y^3-X^3 P_X+15 X Y^2 P_X-12 X Y P_X^2 \\
          && +3 X P_X^3-\frac{37}{12} X^2 Y P_Y+\frac{7 Y^3 P_Y}{3}+\frac{7}{4} X^2 P_X P_Y \\
          && -\frac{25}{4} Y^2 P_X P_Y+\frac{59}{12} Y P_X^2 P_Y-\frac{13}{12} P_X^3 P_Y \\
          && +\frac{13}{8} X Y P_Y^2-\frac{9}{16} X P_X P_Y^2-\frac{Y P_Y^3}{6}-\frac{3}{32} P_X P_Y^3
\end{eqnarray*}
and

\begin{eqnarray*}
a^3 n^3 \chi_3 &=& -\frac{101 X^4 Y}{36}+\frac{200 X^2 Y^3}{9}-\frac{268 Y^5}{45}+\frac{7 X^4 P_X}{4} \\
&& -\frac{329}{6} X^2 Y^2 P_X+\frac{86 Y^4 P_X}{3}+\frac{757}{18} X^2 Y P_X^2-\frac{478}{9} Y^3 P_X^2 \\
&& -\frac{31}{3} X^2 P_X^3+\frac{283}{6} Y^2 P_X^3-\frac{725 Y P_X^4}{36}+\frac{101 P_X^5}{30} \\
&& +\frac{245}{36} X^3
Y P_Y-\frac{58}{3} X Y^3 P_Y-\frac{65}{16} X^3 P_X P_Y+\frac{293}{6} X Y^2 P_X P_Y \\
&&-\frac{145}{4} X Y P_X^2 P_Y+\frac{137}{16} X P_X^3 P_Y-\frac{641}{108}
X^2 Y P_Y^2+\frac{217}{54} Y^3 P_Y^2 \\
&& +\frac{43}{12} X^2 P_X P_Y^2-\frac{89}{8} Y^2 P_X P_Y^2+\frac{223}{27} Y P_X^2 P_Y^2-\frac{293}{144} P_X^3 P_Y^2 \\
&& +\frac{115}{48}
X Y P_Y^3-\frac{41}{24} X P_X P_Y^3-\frac{49 Y P_Y^4}{108}+\frac{83}{192} P_X P_Y^4.
\end{eqnarray*}
We recall that, at each step of normalization, we get new canonical variables on which the normal form depends. However, in view of the result  (\ref{Kzero}), at each order the solution of the canonical equations given by the new Hamiltonian (\ref{Kzero}) 

\ba
\frac{dX}{d \ell} &=& P_X, \quad \frac{d P_X}{d \ell} = - X; \\
\frac{dY}{d \ell}  &=& - \frac34 P_Y, \quad \frac{d P_Y}{d \ell} = 0,\ea
is the same. With a natural choice for the arbitrary constants we can eliminate secular terms and adjust amplitudes so to have the {\it normal mode}

\be \label{NM}
X_{\rm nm} (\ell) =  A_X \cos \ell, \quad
Y_{\rm nm} (\ell) = 0\ee
where $A_X$ is an arbitrary amplitude at our disposal for any needed adjustment. To perform the comparison with the exact epicyclic expansion, we can now apply the rule given by the general transform  (\ref{TNFD}) to the solution on the normal mode to find the series

\ba
{\bv \xi}  (\ell) &=& {\bv \xi}^{(1)} + {\bv \xi}^{(2)} + {\bv \xi}^{(3)} + ...\\
{\bv {p_\xi}} (\ell) &=& {\bv {p_\xi}}^{(1)} + {\bv {p_\xi}}^{(2)} + {\bv {p_\xi}}^{(3)} + ...\ea
Transformations (\ref{T1}--\ref{T4}) will finally provide the series in the initial `physical' variables. The transformations are expressed by the same relations as in the general procedure sketched in (\ref{EHK}) and to second order, in addition to the obvious identities

\ba
 {\bv \xi}^{(1)} &=& (X_{\rm nm}, Y_{\rm nm}),\\
  {\bv {p_\xi}}^{(1)} &=& ({P_X}_{\rm nm}, {P_Y}_{\rm nm}),\ea
  are explicitly given by

\ba
{\bv \xi}^{(2)} &=& L_{\chi_1} {\bv X} = \{{\chi_1},{\bv X}\},\\
{\bv {p_\xi}}^{(2)} &=& L_{\chi_1} {\bv P} = \{{\chi_1},{\bv P}\}\ea
and

\ba
{\bv \xi}^{(3)} &=& L_{\chi_2} {\bv X} + \frac12 L^{2}_{\chi_1} {\bv X} = \{{\chi_2},{\bv X}\} + \frac12 \{{\chi_1},\{{\chi_1},{\bv X}\}\},\\
{\bv {p_\xi}}^{(3)} &=& L_{\chi_2}  {\bv P} + \frac12 L^{2}_{\chi_1} {\bv P}= \{{\chi_2},{\bv P}\}  + \frac12 \{{\chi_1},\{{\chi_1},{\bv P}\}\}.\ea
Finally, by using transformations (\ref{T1},\ref{T2}), we obtain the predictions in terms of the original variables.

To reconstruct in the correct way the whole series we have to properly set arbitrary constants. With the choice made in eqs.(\ref{NM}) we have only to adjust the amplitude of the normal mode. If we write it in the form of a series in the eccentricity

\be
A_X = \sum_{j} a_j e^j,\ee
we can choose the arbitrary coefficients $a_j$ to satisfy the initial conditions. The Kepler orbit we are looking for is such that the periastron is at $t=0$. Therefore a suitable initial condition is $x(0) = - a e, y(0) = 0$: at each order these constraints determine the value of the $a_j, j > 1$ and, as a consequence, the terms in the expansions with the correct coefficients. With the choice $a_1=-a$, we start with the first-order terms

\ba
x^{(1)} &=& -a e \cos \ell ,\\ 
y^{(1)} &=& 2 a e \sin \ell \ea
as in  (\ref{esx1},\ref{esy1}). The constraint at periastron impose the choice $a_2 = 0$ at second order and $a_3 = -a/8 $ at third order, so that we get

\ba
x^{(2)} &=& a e^2 \frac12 \left(\cos 2 \ell -1\right),\\ 
y^{(2)} &=& a e^2 \frac14 \sin 2 \ell \ea
and  

\ba
x^{(3)} &=& a e^3 \frac38 \left(\cos 3 \ell - \cos \ell \right),\\ 
y^{(3)} &=& a e^3 \left(\frac7{24} \sin 3 \ell - \frac38 \sin \ell \right). \ea
Comparing with (\ref{esx1},\ref{esy1}) we see that the predictions offered by the normal form approach are exact up to order $e^3$. Going even farther, we get, at fourth order, $a_4 = 0$ and

\ba
x^{(4)} &=& a e^4 \left(\frac{67}{192} \cos 4 \ell - \frac13 \cos 2 \ell - \frac{1}{64}\right),\\ 
y^{(4)} &=& a e^4 \left(\frac{29}{96} \sin 4 \ell - \frac5{12} \sin 2 \ell \right),  \ea
which agrees with results obtained by patiently solving the equations of motion (see e.g. Brouwer and Clemence, 1961). 
	
\section{Normalization of the perturbed Kepler problem}

In the previous section we have validated the normal form approach by exploiting a high order (N=4) normalization of the epicyclic approximation of the Kepler problem. We can then rely on a powerful method of retrieving information on the dynamics of the whole class of systems given by autonomous perturbations of the Kepler problem.

Referring to the perturbation parameter introduced in (\ref{Omegap}) and (\ref{NF}), two important examples are: the `Schwarzschild' correction mimicking the relativistic precession given by

\be\label{Schc}
\varepsilon_{\rm S} \equiv \frac{3r_{\rm S}}{2 a},\quad r_{\rm S} =\frac{2GM}{c^2} \ee
and a gravitational quadrupole correction given by

\be\label{Solc}
\varepsilon_{\rm Q} \equiv \frac32 J_2 
\left(\frac{R}{a}\right)^2\ee
where $R$ is the radius of the central body. In both cases the parameters $r_s$ and $J_2 \times R^2$ are `small' in these units  (Pucacco et al., 2010). 

Another potential of great relevance is the so called {\it isochrone}  (Boccaletti and Pucacco, 1999) that in this framework can be approximated by

\be\label{isoc}
\varepsilon_{\rm I} \equiv - \frac{b}{a},\ee
where $b$ is a length-scale assumed small with respect to $a$.

It happens that, in general, the normal form is more complex than in the case of the Kepler problem. In particular, it is no more possible to eliminate all terms with the canonical transformation. We discuss below the origin and the implications of this phenomenon. 
In our derivation we stop at the first steps of normalization procedure to avoid cumbersome formulas. However, this is still a very useful starting point for applications; in fact, in the spirit of perturbation theory and in view of the smallness of all coupling constants introduced above, terms in the solutions coming from these perturbations can be linearly superposed to terms giving the `unperturbed' solution of the previous section. 

The order at which both series have to be truncated depends on the required precision and on the relative size of the perturbation terms. For example, in the case of the orbits of LISA ($a = 1 UA, \; e = 0.01$), if one is interested in investigating the Schwarzschild precession due to the Solar field ($\varepsilon_{\rm S} \simeq 10^{-8}$), in principle the Kepler expansion should be truncated at order $e^4$: however, in practice, in view of the differential motion on the members of the `constellation' that gives the measurable effect, one can truncate at order $e^3$ (Pucacco et al., 2010). In any case, we see that the greatest part of the `job' is done by the Kepler expansion and one step of normalization of the additional perturbation is enough for our purposes.

Therefore, we limit ourselves to writing the explicit form of the first non-vanishing terms of the normal form and of the generating function. They are:

\be \label{K2}
{\cal K}_{2} = -\frac{\varepsilon}{a^2} \left[\frac{17}{4} (P_X^2 + X^2)^2 + 30 Y^2  (P_X^2 + X^2) + 10 Y^4 \right]\ee 
and
 
\begin{eqnarray*}
\frac{a}{\varepsilon} \chi_1 &=& - 5 X^2 Y + \frac{40 Y^3}{9} + \frac{10}3 X^2 P_X - 10 Y^2 P_X + 5 Y P_X^2 \\
          && -\frac{10 P_X^3}{9} + 5 X Y P_Y - \frac{25}{4} X P_X P_Y - \frac{40 Y P_Y^2}{9} + \frac{25}{4} P_X P_Y^2.
\end{eqnarray*}
Comparing this last expression with the corresponding one in the Keplerian case (\ref{chi1}) we can see the slightly different approach in the two perturbation settings: here the perturbation parameter appears explicitly in the generating function (and in the normal form) whereas in the Keplerian case the role of the perturbation parameter is played by $e$ and is hidden into the coordinates. 

\begin{figure}[h!]
\centering
\includegraphics[width=0.49\columnwidth]{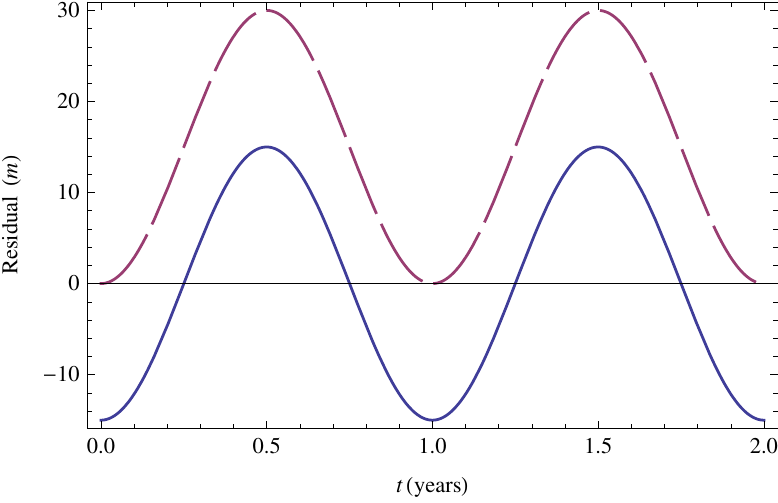}
\includegraphics[width=0.49\columnwidth]{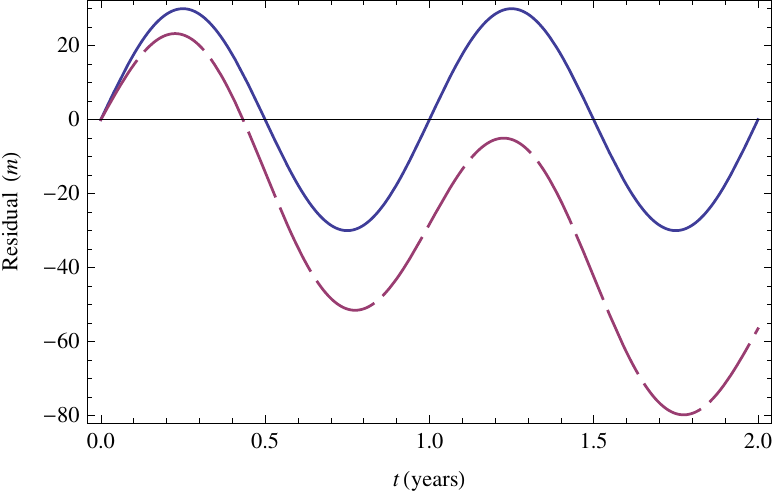}
\caption{Approximations of the Schwarzschild precession with the post-epicyclic formulas in the Hill system: the $x$ coordinate is on the left, the $y$ coordinate is on the right. The continuous lines represent the residuals from the exact numerical evaluation obtained with the second-order expressions (\ref{2sx}--\ref{2sy}) whereas the dashed lines are obtained with the first order approach of  Pucacco et al. (2010). The orbital parameters are $a = 1 UA, \; e = 0.01$.}
\label{finale}
\end{figure}

Afterwards, proceeding as in the previous section, we obtain the second-order solution to the equations of motion representing the effect of the field perturbation:

\ba
x^{(\varepsilon)} &=& a e \varepsilon \left(-\frac76 -\cos \omega_1 t  + \frac{13}6 \cos 2\omega_1 t \right),
\label{2sx}\\ 
y^{(\varepsilon)} &=& 2a \varepsilon \left[(1 + e) \sin \omega_1 t + \frac1{24}e \sin 2 \omega_1 t \right]. \label{2sy} \ea
This solution is one order (in $e$ and $\varepsilon$) beyond those found in the description of the precessional effects to which is sensitive the LISA constellation (Pucacco et al., 2010); however, in that work, the first order (in $\varepsilon$) terms have a more complex structure because that solution was computed in a Hill frame rotating at the fixed value $n$ of the unperturbed Keplerian orbit. The additional constant part in $x(t)$ appearing in Pucacco et al. (2010), $-\varepsilon a/3$, accounts for the change of radius of the reference orbit due to the slight change in the force. Here this term disappears because the different angular velocity of the Hill frame gives a centrifugal force which exactly balances the total active force due to the Keplerian + perturbing field. A constant term appears only at second order. In Fig.\ref{finale} we show an example of the use of the two approximations.

Another important difference in these solutions is the presence of the new frequency $\omega_1$. This occurs precisely because the second order term (\ref{K2}) in the normal form is non-vanishing. If we write the generalization of the normal mode (\ref{NM}) in the form

\be\label{NM2}
X_{\rm nm} (t) = \sqrt{2 J} \cos \omega_1 t, \quad Y_{\rm nm} (t) = 0,\ee
where $J$ is an action-like variable, its frequency can be computed from the Hamiltonian as

\be
\omega_1 = \frac{\partial {\cal K}}{\partial J},\ee
where the derivative is evaluated {\it on} the normal mode (\ref{NM2}). It is easy to check that

\be
\omega_1 = \omega \left(1 - \frac{17 \varepsilon}{2 a^2} J \right).\ee
In practice, this result has negligible consequence on the predicted amount of these small effects: the advance of periastron is still approximated with high accuracy by the formula

\be
\Delta \psi = 2 \pi \frac{\Omega - \omega}{ \omega} \simeq 2 \pi \varepsilon \;\; [{\rm rad / orbit}] \ee
because the extra correction is of second order in the perturbation parameter. However, the result is important from a formal point of view, since it accounts for the dependence of frequency on the amplitude of oscillation. This remark shows us the relevance of the non-vanishing second (and higher) order terms in the normal form: only in this way this general phenomenon may happen. In the degenerate case of the exact Kepler problem (and in the trivial case of the harmonic oscillator) both frequencies $\Omega$ and $\omega$ are constant and coincide with the mean motion, so that $\Delta \psi=0$ and all orbits are closed. In this perturbative setting this happens only if {\it the normal form coincides with the quadratic term at every order}.

\section{Conclusions}

The results presented in this work provide a comprehensive setting to understand some stimulating but unsettled results obtained in previous studies (Gurfil and Kasdin, 2003; Kasdin et al., 2005; Durandhar et al., 2005; Rajesh Nayak et al., 2006; Sweetser, 2005; Pucacco et al., 2010).  

We have here tried to address the natural question about the limits of validity of those results. We adopted the approach usually followed in exploiting approximate invariants constructed in a high order perturbation theory (Giorgilli and Locatelli, 2006) and we have checked, in the specific problem of the perturbed Keplerian motion, how the power series representing the approximate solution can be used to imitate the exact solution. Overall, the approach based on the normal form extends in many ways the results of works based on the Hamilton-Jacobi equation and perturbations thereof (Kasdin et al., 2005) and on the simple direct analysis of the non-linear equations of motion (Rajesh Nayak et al., 2006; Pucacco et al., 2010).

In the framework based on the Hill reference system, we can state that a perturbative approach makes sense if the expansion of the total potential is performed up to a degree such that the series representing the post epicyclic approximation of the unperturbed orbits and the amount of the perturbation give errors of comparable size. These results also provide answers to the questions raised by Sweetser (2005) concerning the relevance of the post-epicyclic series in satellite problems. In the study of satellite formations on orbits with small eccentricity this approximation is extremely good (Pucacco et al., 2010) and can be also profitably used in the coorbital problem (Gurfil and Kasdin, 2003) and in the so called `tethered' system (Sidorenko and Celletti, 2010). 

The generalization to the 3-dimensional case can be performed with limited effort. At the linear level, the oscillation in the $z$ direction is decoupled and proceeds with its own proper frequency. In the presence of a perturbation, the non linear coupling is managed through the normal form with standard methods (Giorgilli, 2002). Non-autonomous perturbations can be considered by extending the phase-space to include the time variable. In general, in 3 dimensions, the normalizing transformation no longer provides an integrable normal form. Usually, the improvement consists only in lowering by one the dimensionality of the problem. However, the investigation of existence and stability of normal modes proceeds in much the same way as above. 

\section*{Acknowledgments}
This work is supported by INFN -- Sezione di Roma Tor Vergata. I thank Massimo Bassan for a careful reading of the manuscript and Fabrizio De Marchi for stimulating discussions.

\end{document}